\documentclass[utf8]{FrontiersinVancouver}

\usepackage{url,hyperref,microtype,subcaption}
\usepackage[switch,columnwise]{lineno}
\usepackage[onehalfspacing]{setspace}
\usepackage{braket}
\usepackage{tikz}
\usetikzlibrary{positioning,shapes,arrows}
\usepackage{array}
\usepackage{booktabs}
\newcolumntype{L}[1]{>{\raggedright\arraybackslash}p{#1}}
\makeatletter
\renewcommand\@biblabel[1]{[#1]}
\makeatother
\def\keyFont{\fontsize{8}{11}\helveticabold }
\def\firstAuthorLast{Cacciapuoti {et~al.}}
\def\Authors{Angela Sara Cacciapuoti\,$^{1,*}$, Jessica Illiano\,$^{1}$, Caterina De Risi\,$^{1}$ and Marcello Caleffi\,$^{1}$}
\def\Address{%
$^{1}$\href{https://qnattynet.quantuminternet.it/}{www.QuantumInternet.it} research group, University of Naples Federico II, Naples, 80125, Italy}
\def\corrAuthor{Angela Sara Cacciapuoti}
\def\corrEmail{angelasara.cacciapuoti@unina.it}

\begin{document}
\onecolumn
\firstpage{1}

\title[Beyond Quantum Messages: Toward Quantum-Native Network Functioning]{Beyond Quantum Messages: Toward Quantum-Native Network Functioning}

\author[\firstAuthorLast]{\Authors}
\address{}
\correspondance{}
\extraAuth{}

\maketitle

\begin{abstract}

\section{}

The Quantum Internet is commonly envisioned as a network that transports quantum states and distributes entanglement as the key resource underpinning distributed quantum applications. This view is necessary but incomplete if quantum phenomena are confined to what the network carries or manages, while network functionalities remain grounded predominantly in classical abstractions. This Perspective advances a broader thesis: quantum phenomena can also participate directly in the realization of network functionalities. We first establish the architectural baseline by introducing entanglement as a stateful network resource whose operational condition is non-locally distributed across remote nodes, evolves through manipulation or consumption, and degrades over time through decoherence. Its evolving state must therefore remain visible to the functionalities acting on it. We then move beyond resource visibility and use two mechanisms as proof points of a broader quantum-native functional design space: entanglement generation during propagation through coherent spatial-path superposition, and operational connectivity induced by shared entanglement that can depart from the physical graph. These examples show that paths, neighbors, or participating nodes may become operationally relevant in ways determined by quantum coherence and shared entanglement, rather than by physical connectivity alone. We identify Quantum Addressing as a transversal abstraction capable of representing such relations while complementing, rather than replacing, classical addressing and signaling. The resulting perspective broadens the Quantum Internet from transporting quantum information and distributing entanglement to exploiting quantum phenomena in network functioning, opening a coupled frontier spanning routing, forwarding, control, and functionality placement.

\tiny
\keyFont{\section{Keywords:}
quantum path, entanglement, quantum addressing, Quantum Internet, quantum-native functioning, ERC CoG QNattyNet}
\end{abstract}

\section{Beyond Quantum Messages}
\label{sec:01}

We start this Perspective from a distinction that broadens the design space of quantum communications and networking: the Quantum Internet should not be understood only in terms of the quantum information it carries or the entanglement it distributes. The complementary question is: \textit{what becomes possible when quantumness is admitted not only into what the network carries or manages, but also into how the network functions?}

This question can be approached through progressively richer, but tightly connected, roles of quantumness in network operation. At the most basic level, a quantum state --including an entangled state-- can be regarded as the object that the network is required to establish, transport, or deliver among remote nodes. From this viewpoint, the state is treated through a message abstraction: the networking problem is formulated in terms of establishing and delivering the target quantum state, while its control logic used to route, schedule, or coordinate that process may remain predominantly classical \cite{KozWehVan-22}.

Entanglement, however, goes beyond the message abstraction once it is shared among network nodes. It becomes a stateful network resource: its operational state is non-locally distributed, is locally transformed or depleted through use, and degrades over time through decoherence. More importantly, the availability of entanglement changes what the network can operationally do. It can induce connectivity relations that depart from the underlying physical graph, enabling network functionalities that exploit such shared quantum correlation. For instance, nodes that share entanglement can become operationally connected even when they are not physically adjacent, enabling routing and forwarding mechanisms that operate over entanglement-defined relations rather than exclusively over physical next-hop adjacency \cite{CalCac-25}, as further discussed in Section \ref{sec:03.2}. More generally, such relations make available network primitives whose operation depends directly on the shared quantum resource. This operational connectivity relies on entanglement established through the underlying physical network rather than on additional deployment of physical links. Thus, entanglement is not only something to be delivered: its evolving condition becomes part of the network state that functionalities must be able to observe and act upon \cite{IllCalMan-22,CalCac-25}. Once entanglement can itself define the operational connectivity and enable network primitives, quantumness need no longer remain confined to the object being managed. It can also become part of the mechanisms through which network functionalities are realized.

We refer to this broader design direction as quantum-native network functioning. Here, quantum phenomena and quantum-mechanical laws are exploited not only in the message being carried or in the entanglement resource being managed, but in the realization of network functioning itself. In this regime, superposition, quantum interference, entanglement, or coherent control become constitutive elements of the mechanisms through which networking functions are performed \cite{CacCal-26,RamPirDur-21,CalCac-25,CacCalIll-25}.

Much of the current architectural landscape for the Quantum Internet naturally focuses on transporting quantum states and on generating, distributing, and managing entanglement as a network resource \cite{AguLopLop-19,Weh-19,VanRyoBen-22}, while addressing, routing, resource coordination, and control remain predominantly grounded in classical representations and signaling. This is a coherent and productive strategy, particularly suited to near-term deployment, and classical mechanisms remain indispensable. The thesis advanced here is therefore additive rather than substitutive: classical abstractions need not be discarded, but they need not exhaust the functional design space either. This is also the architectural distinction we have emphasized elsewhere \cite{CacCalIll-25}: quantum-native abstractions augment rather than replace classical control and signaling.

\begin{figure*}[htbp]
\centering
\includegraphics[width=\textwidth]{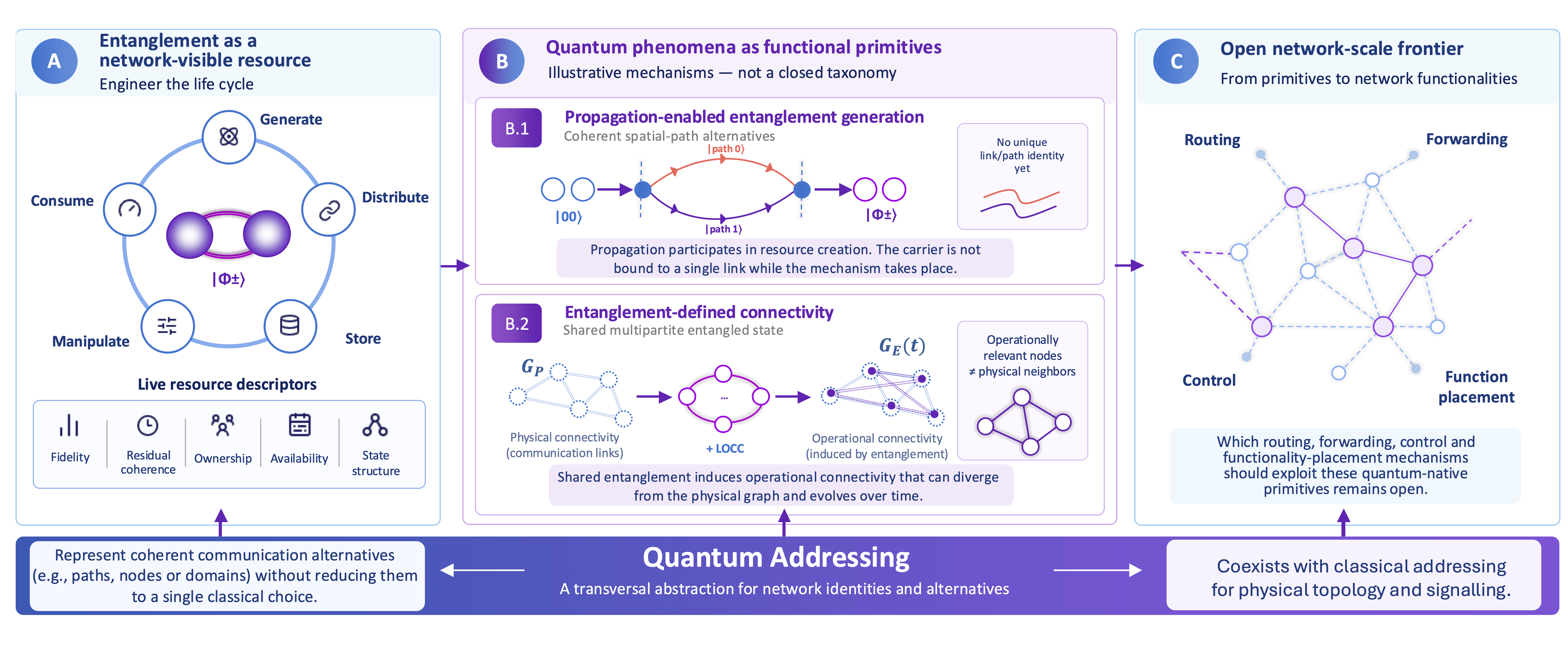}
\caption{From network-visible entanglement to quantum-native network functioning. Entanglement goes beyond the message abstraction as a stateful resource whose evolving condition must remain visible to the network. Beyond visibility, quantum phenomena can participate directly in network functioning, as illustrated by propagation-enabled entanglement generation (panel B.1) and entanglement-defined connectivity (panel B.2). Note that the physical connectivity captured by the physical graph $G_P$ is necessary to distribute link-level entanglement. Through the entanglement relations established through entanglement distribution over $G_P$ and its manipulation via LOCC (Local Operation and Classical Communications) give the entanglement-defined operational graph $G_E(t)$, which changes over time and can be different from the physical graph, i.e., $G_E(t) \neq G_P$.} These mechanisms motivate complementary quantum-grounded abstractions and open a coupled frontier spanning routing, forwarding, control, and functionality placement.
\label{fig:beyond-messages}
\end{figure*}

Emerging results already show that quantumness can enter network functionalities themselves \cite{RamPirDur-21,CalCac-25,CacCalIll-25}. Yet the broader functional design space enabled by this shift remains comparatively unexplored. We explore it through two concrete mechanisms, used as proof points rather than as a complete architecture: entanglement generated during propagation through coherent spatial-path superposition, and operational connectivity induced by shared entanglement that can depart from the physical graph. Together, they show that the relevant paths, nodes, or connectivity relations seen by a network function may themselves depend on quantum coherence or entanglement.
This creates a representational problem before it creates a routing problem: the network needs abstractions capable of expressing such quantum-defined relations while classical identities, physical graph, and signaling remain available. Quantum Addressing provides one such transversal abstraction. As pictorially depicted in Fig.~\ref{fig:beyond-messages}, beyond representation, these mechanisms expose a coupled frontier spanning routing, forwarding, control, and functionality placement, which the remainder of this Perspective sets out to delineate.

\section{Entanglement as a Network-Visible Resource}
\label{sec:02}

Once entanglement is shared among network nodes, its statefulness creates a distinctive architectural problem. The non-local resource can be transformed or consumed by operations performed locally at its participating nodes, and degrades over time through decoherence. Any network-level representation of that resource must therefore remain meaningfully associated with the particular entanglement instance to which it refers as the resource evolves. If the resource is consumed, reconfigured, lost, or replaced, a representation that is no longer aligned with it may lead a network functionality to reason about a resource that no longer exists in the same operational form. 

Network visibility therefore has at least two dimensions: \textit{keeping a representation causally aligned with an evolving distributed resource, and ensuring that the representation preserves the properties on which the intended network functionality depends}. Note that network visibility does not require exposing the complete description of the shared quantum resource state to every network functionality acting on it. Differently, the information that must be exposed depends on the operation to be performed and on the resource properties relevant to its execution.

A binary availability flag is clearly insufficient. Knowing only that entanglement exists between a set of nodes says little about whether it can support a specific operation at a particular time. Depending on the functionality, relevant descriptors may include the participating nodes, fidelity or another task-relevant quality measure, residual coherence budget, availability, ownership, and the structure of the shared state for the multipartite-entanglement regime. Different functionalities may require different views of the same resource: a scheduler, a purification procedure, or a service-selection mechanism need not rely on the same descriptors or on the same degree of state knowledge.

The architectural issue, however, cannot be reduced to maintaining richer classical descriptors. \textit{As long as the representation is maintained separately from the quantum resource, every transformation, consumption, loss, or reconfiguration of that resource creates an association and causal-consistency obligation}: the classical view must be reconciled with the new resource state before subsequent network decisions can safely depend on it. 

Additional classical signaling can perform this reconciliation, but it does not remove the underlying separation that makes reconciliation necessary in the first place. As the number of distributed resources, participating nodes, concurrent operations, and administrative domains grows, the scope over which correctness depends on such reconciliation becomes itself a scalability concern \cite{CacCal-26,CacCalIll-25}.

A second, equally important aspect is the semantic limitation of a classical-only representation. A classical description can encode properties of a quantum resource, and can even describe that several alternatives coexist coherently, but it cannot itself preserve those alternatives as a quantum state or support coherent operations on them. Hence, even an ideally reconciled classical view is not operationally equivalent to a quantum-grounded abstraction. This distinction becomes essential whenever a network function must not merely know about a quantum relation, but preserve or act on its quantumness as part of the function itself.

Network visibility should therefore not be understood as classical monitoring of a quantum resource. It calls for complementary abstractions: classical descriptors and signaling remain indispensable for physical-graph visibility and the communication of measurement outcomes, while quantum-grounded representations are required to reduce the architectural dependence on separately reconciled classical state and to preserve quantum relations as operational control variables. 

The issue is thus not whether classical signaling can operate a quantum network, but whether classical representation alone should define the abstraction boundary through which the network reasons about an evolving quantum resource. The next section makes this distinction concrete. Spatial-path superposition and entanglement-defined connectivity provide two cases in which the quantum character of the relevant network relation is not merely something to describe: it participates in what the network function can do.

\section{When Quantumness Enters Network Functioning}
\label{sec:03}

Section~\ref{sec:02} showed that making entanglement network-visible already raises the need for abstractions grounded in the evolving quantum resource. We now take the next step and ask what changes when quantum phenomena participate in the mechanism realizing a network function itself. We develop this point through two proof cases in which quantumness changes, respectively, how a resource is created and what connectivity a functionality can exploit.

\subsection{Beyond Transport: Quantumness in Propagation}
\label{sec:03.1}

Conventional entanglement-distribution models conceptually separate resource creation from distribution. Entanglement is first established by a source and then distributed by propagating the corresponding quantum systems over a communication link selected by the network. At this level, the physical evolution of the transmitted quantum systems is quantum, but the propagation alternative itself is resolved classically: a definite link is selected before transmission takes place.
Quantumness can instead be lifted to the propagation process itself. Spatial-path superposition exploits a quantum control degree of freedom to coherently control the use of two or more spatially distinct communication links \cite{EblSalChi-18,ChiKri-18,SkPelCal-26,PelSkCal-26,CacPelCal-26,ChaShiMig-26}. Rather than applying one classically selected channel map, the path-control degree of freedom enables the alternative path-dependent channel evolutions to act coherently on the target system. 

The resulting interference terms have no counterpart in an incoherent classical mixture of the same propagation alternatives, and change the functional role of propagation in two distinct ways. First, propagation can participate directly in resource creation. Coherently exploiting alternative channel evolutions can deterministically transform initially separable states into bipartite or multipartite entangled states, so that entanglement is generated during propagation itself, rather than generated first and subsequently transported \cite{PelSkCal-26}. Generation and transport therefore cease to be necessarily distinct stages of the entanglement lifecycle.

Second, the same interference structure can reshape how noise affects the transmitted state. Rather than combating noise only after a definite noisy channel has acted, coherent control of alternative channel evolutions can make their interference part of the protection mechanism itself. Across different classes of noise models, this can increase the fidelity of the delivered state, suppress otherwise detrimental noise contributions, or activate communication properties unavailable from the individual alternatives \cite{CalCac-20,KouCacCal-21,RamPirDur-21,CacPelCal-26}. This is not quantum error correction in the coding-theoretic sense; it is a genuinely quantum form of noise mitigation or protection arising from coherence at the propagation level.

The common lesson is that propagation is no longer a passive mechanism for carrying a pre-existing quantum resource. Its quantum character can contribute to creating that resource and to protecting the state while it propagates. More fundamentally, spatial-path superposition makes concrete the limitation of the classical representations discussed in Section~\ref{sec:01}. A quantum delivery process coordinated exclusively through a classical, topology-grounded abstraction such as IP-like addressing cannot natively capture the situation in which, while the primitive is active, the quantum message is not associated with one definite propagation trajectory. Its spatial degree of freedom is instead coherently delocalized across alternative links.
The limitation is therefore representational: a classical abstraction grounded in resolved topological information cannot encode the message location itself as a coherent quantum variable. 

Exploiting such a primitive at network level consequently calls for a quantum-grounded abstraction capable of preserving this unresolved quantum relation rather than projecting it onto a single topological alternative.
This, in turn, raises a control question: how should such quantum-grounded propagation alternatives be exposed to routing and control so that they can be identified and acted upon without being classically resolved before the functionality has exploited their coherence? Classical signaling remains necessary, but it must coexist with abstractions able to carry the quantum semantics of the propagation alternatives themselves.

\subsection{Entanglement as a Connectivity Primitive}
\label{sec:03.2}

A second, complementary example shows how entanglement can participate in network functioning not only as a resource to be managed, but by defining the connectivity on which a network functionality operates. Let \(G_P\) denote the physical network graph, determined by the physical communication channels interconnecting the nodes. For a given quantum network functionality, however, physical adjacency need not coincide with operational, entanglement-induced adjacency. Two nodes that are not directly connected in \(G_P\) may share entanglement and thereby become operationally connected for a functionality that can exploit that resource \cite{CheIllCac-24,MazCalCac-24-1}. We denote by \(G_E(t)\) the corresponding time-dependent entanglement-defined operational graph. In general,
\[
G_E(t) \neq G_P.
\]

This distinction directly instantiates the transition introduced in Section~\ref{sec:01}: entanglement is no longer only a stateful resource whose condition must be exposed to the network. Its quantum correlations can determine which network relations are operationally available and, consequently, which functionalities can be realized. Unlike $G_P$, $G_E(t)$ is not fixed by the deployment of physical channels. It depends on the entanglement shared at time $t$ and on whether that resource remains usable for the operation under consideration, according, for instance, to its fidelity, residual coherence time, and the structure of the shared state. As entanglement is generated, consumed, degraded, or reconfigured, $G_E(t)$ changes accordingly while $G_P$ may remain unchanged.
Entanglement can therefore make connectivity itself programmable \cite{MazPelCac-26}. By manipulating the shared quantum resource, the network can reconfigure the operational relations available to a functionality without modifying the underlying physical infrastructure. Which nodes become operationally connected, which operations they can perform, and which service can consequently be realized are jointly determined by the shared quantum state and by the operations admissible on it. Hence, the graph relevant to network functioning is not only the physical graph --which remains the necessary substrate for establishing and distributing quantum resources-- but also the one emerging dynamically from the quantum resources currently available to the network.

This exposes a representational limitation complementary to the one identified for spatial-path superposition. There, a topology-grounded classical representation cannot capture a propagation relation that remains coherently unresolved. Here, physical graph alone cannot capture operational relations induced by entanglement between physically remote nodes. In both cases, the network relation relevant to the functionality is determined, at least in part, by a quantum state rather than solely by the physical graph. Classical topology-driven abstractions, such as IP-like addressing and signaling, therefore remain necessary, but they are no longer sufficient to represent all the relations on which network functionalities may operate.

Taken together, spatial-path superposition and entanglement-defined connectivity expose the same architectural requirement from two different directions. The former requires representing a propagation relation that remains quantum while the functionality is executed; the latter requires representing operational relations that do not coincide with physical adjacency. In both cases, the network relation on which the functionality operates cannot be fully captured by resolved classical identities and physical-topology information alone. This motivates complementary quantum-grounded abstractions capable of supporting network operation over such quantum-defined relations, while coexisting with the classical identities and signaling that remain indispensable. It also foreshadows a broader issue: a functionality may be logically associated with a node or service while the mechanism realizing it spans multiple links, remote nodes, or a non-local shared state.

The next section considers Quantum Addressing as one such transversal abstraction.

\section{Quantum Addressing as a Transversal Abstraction}
\label{sec:04}

The two examples in Section~\ref{sec:03} expose a common representational requirement: the relation on which the functionality operates cannot be fully expressed if network identity has already been reduced to a single classical, topology-grounded alternative.

To meet this requirement while preserving the classical identities and signaling needed for physical-topology operation, Quantum Addressing provides one such identity abstraction \cite{CacIllCal-23,CalCac-25}: a node identity can be encoded as a quantum address, while a set of candidate identities can be represented coherently rather than as a classically enumerated set from which one alternative has already been selected. Quantum operations can then act on this representation without first converting the relevant identity relation into a resolved classical choice.

More in detail, Quantum Addressing is a \textit{transversal} abstraction because it separates two questions. The first is \textit{who} may participate in a network operation, which quantum addressing answers through a coherent representation of the candidate identities. The second is \emph{what} operation is to be performed, which remains with the functionality consuming that representation. This makes the identity abstraction transversal, rather than one tied to a specific network functionality or protocol layer. Different network functionalities can therefore exploit the same identity abstraction according to their respective operational requirements.

In the propagation example, Quantum Addressing does not itself create the spatial superposition or encode the complete channel dynamics. Rather, it can associate coherent propagation alternatives with the corresponding endpoint, intermediate-node, or candidate-neighbor identities. In the connectivity example, it can represent the identities of nodes participating in operational relations induced by $G_E(t)$ independently of physical adjacency. Entanglement-aware routing \cite{CalCac-25} and the service intent of the protocol suite in \cite{CacCal-26}, which identifies participating nodes through quantum addresses, illustrate this transversality at different architectural levels, rather than as obligatory components of a single architecture.

Existing results already demonstrate quantum-native routing and forwarding over entanglement-defined topologies using Quantum Addressing \cite{CalCac-25}. They do not, however, address the distinct case in which the propagation path itself remains in coherent superposition. Quantum Addressing can therefore provide an enabling identity abstraction for extending network operation to such propagation regimes, without itself encoding the path superposition or its channel dynamics. Quantum Addressing should therefore be understood as an enabling abstraction for reasoning about packets or operations associated with coherent propagation and connectivity alternatives. 

In other words, Quantum Addressing provides a network-level handle on identities associated with quantum-defined relations, preserving coherent identity alternatives where required. Once such identities can be exposed and manipulated without forcing the underlying network relation into a purely topology-grounded classical representation, the architectural question shifts from representation to use: how should routing and forwarding, hybrid control, and protocol organization and composition operate over them at network scale? This is the frontier considered next.

\section{Open Network-Scale Frontier}
\label{sec:05}

The mechanisms discussed in this Perspective belong to a broader quantum-native networking direction in which quantum phenomena participate directly in network functionalities \cite{PirDur-19,RamPirDur-21,CacCalIll-25,PirMunDur-25,CacCal-26}. 

Quantum Addressing, quantum-native routing, and generalized quantum forwarding provide concrete examples in which superposition and quantum-state processing are exploited within network operation itself \cite{CalCac-25,AbaCubMai-25,CacCalIll-25}, while dynamic composition provides a quantum-native principle for protocol organization \cite{CacCal-26}. The direction is not confined to this layer: at the node level, recent work addresses the programmability of quantum network nodes, from an operating system executing networked applications on quantum nodes \cite{DelIulVan-25} to an instruction-set architecture for NV-center repeater nodes that includes coherent control of a quantum register \cite{KumCicBas-26}. The network-scale frontier is therefore how different quantum-native mechanisms can coexist, interact with indispensable classical mechanisms, and compose into a scalable and operable architecture. Three aspects are particularly relevant. Table~\ref{tab:network_scale_frontier} summarizes the resulting network-scale frontier along these three coupled dimensions.

\begin{table}[t]
\centering
\caption{Network-scale frontier for quantum-native functioning.}
\label{tab:network_scale_frontier}

{
\renewcommand{\arraystretch}{1.25}
\setlength{\tabcolsep}{3pt}

\begin{tabular}{
@{}
>{\raggedright\arraybackslash\bfseries}p{0.27\columnwidth}
>{\raggedright\arraybackslash}p{0.69\columnwidth}
@{}
}
\hline
Dimension & \textbf{Network-scale frontier} \\
\hline

Routing \& forwarding
&
How to support route computation and forwarding across physical-graph,
entanglement-defined, and coherently superposed propagation regimes, and how
these regimes should interact without prematurely reducing quantum-defined
relations to a classical path.
\\[0.9em]

Hybrid control
&
How to determine and manage the quantum--classical control boundary: which
network relations must remain quantum-grounded, which can be represented
classically, and how the two remain consistent as quantum resources evolve.
\\[0.9em]

Protocol organization \& composition
&
How to instantiate and validate dynamic composition across a broader
repertoire of quantum-native primitives while preserving the quantum features
on which their functional advantage depends.
\\

\hline
\end{tabular}
}
\end{table}

\textit{Routing and forwarding}. The classical distinction between route computation and forwarding remains useful also for entanglement routing, although both phases acquire different semantics \cite{AbaCubMai-25}. Physical-topology information remains necessary for computing and provisioning the elementary links through which entanglement can be generated. At the same time, quantum-native routing can operate over the graph activated by already available entanglement, with route selection, depending on the routing regime, informed by entanglement availability and by quantum operations through which end-to-end connectivity can be established \cite{CalCac-25,CacCalIll-25}. Forwarding -- a.k.a. generalized forwarding -- then realizes the selected route through operations such as entanglement generation, swapping, or purification rather than through classical match-and-forward. The propagation mechanism of Section~\ref{sec:03.1} introduces an additional routing dimension: the communication path itself may remain in coherent superposition while the functionality is executed. The open question is how such propagation-level quantum alternatives should be exposed to quantum-native routing and forwarding without being prematurely reduced to a single classical path. 

\textit{Hybrid control}. Classical signaling remains necessary for coordination, measurement outcomes, physical-resource management, and interoperability, while quantum representations are useful when the relevant network relation itself must retain quantum properties. The resulting control problem is therefore intrinsically hybrid. A key open question is how classical and quantum control information should interoperate: which control variables or network relations must remain represented and manipulable in quantum form, which can be resolved into classical state, and through which interfaces the two representations remain mutually consistent as resources evolve.

\textit{Protocol organization and composition}. The non-local and stateful nature of entanglement already challenges rigid protocol layering. The dynamic-composition framework in \cite{CacCal-26} addresses this problem by replacing fixed vertical organization with the runtime composition of atomic network functionalities according to the purpose of the service, the capabilities of the nodes, and the current state of quantum resources. Quantum Addressing is embedded in this model both in the service intent, to identify the participating nodes, and in the quantum forwarding tables, to associate evolving entanglement resources with the corresponding network identities. The resulting resource state then enters the local composition logic. The mechanisms considered here reinforce this architectural choice: primitives such as propagation-enabled entanglement generation or entanglement-defined connectivity cut across conventional functional boundaries and are naturally treated as composable network primitives rather than assigned a priori to fixed layers. The remaining challenge is to instantiate and validate this compositional organization across a broader repertoire of quantum-native primitives while preserving their quantum functional advantage.

Taken together, these directions suggest that quantum-native networking should not be understood as replacing classical networking mechanisms, but as extending the functional design space through quantum abstractions and primitives where they provide an operational advantage. The resulting research challenge is to determine how such primitives can be integrated with classical signaling and control, and dynamically composed at network scale without losing the quantum features that underpin their functional advantage. 

In this sense, the Quantum Internet design problem extends beyond carrying quantum information and distributing entanglement to exploiting quantum phenomena in the realization of network functioning itself.


\section*{Funding}

This work has been funded by the European Union under Horizon Europe ERC-CoG grant QNattyNet, n.101169850. Views and opinions expressed are however those of the author(s) only and do not necessarily reflect those of the European Union or the European Research Council Executive Agency. Neither the European Union nor the granting authority can be held responsible for them.

\bibliographystyle{Frontiers-Vancouver}
\bibliography{biblio}

\end{document}